\documentclass[journal=jpccck,manuscript=article]{achemso}
\usepackage[T1]{fontenc} 
\usepackage{amsmath}
\usepackage{amssymb}
\usepackage{natbib}
\usepackage[svgnames]{xcolor}
\usepackage[colorlinks=true,citecolor=blue,linkcolor=blue,citebordercolor=white,linkbordercolor=white]{hyperref}
\usepackage{soul}
\usepackage{gensymb}

\def\eq#1{{\textcolor{blue}{eq} \ref{#1}}}

\def\Eq#1{{\textcolor{blue}{Equation} \ref{#1}}}

\def\onlinecite#1{{ref \citenum{#1}}}

\def\onlinecites#1{{refs \citenum{#1}}}

\def\fig#1{{\textcolor{blue}{Figure} \ref{#1}}}

\def\vec#1{{\bf{#1}}}

\author{S.V. Novikov}
\email{novikov@elchem.ac.ru}
\phone{+7 495 952 24 28}
\fax{+7 495 952 53 08}
\affiliation[Frumkin Institute]
{A.N. Frumkin Institute of
Physical Chemistry and Electrochemistry, Leninsky prosp. 31,
119071 Moscow, Russia}
\alsoaffiliation[HSE]
{National Research University Higher School of Economics, Myasnitskaya Ulitsa 20, Moscow 101000, Russia}

\title[Recombination]{Enhanced Bimolecular Recombination of Charge Carriers in Amorphous Organic Semiconductors: Overcoming the Langevin Limit}

\SectionNumbersOn

\begin{document}

\begin{tocentry}
\includegraphics[width=1.75in]{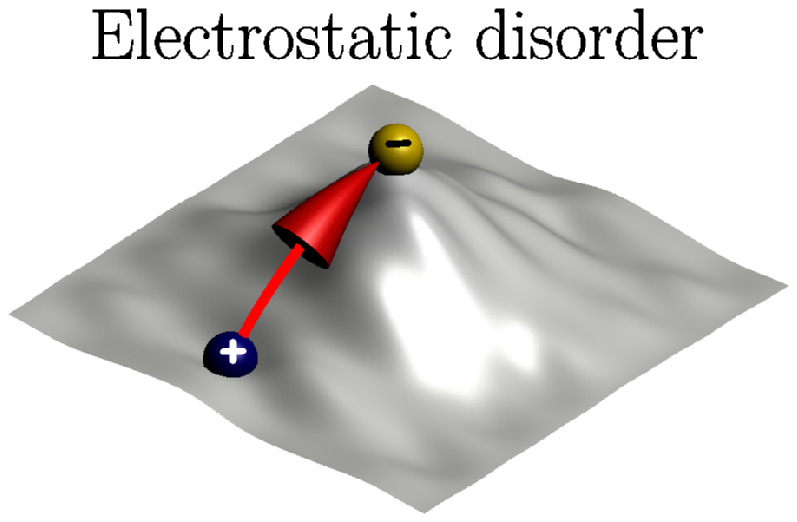}
\includegraphics[width=1.75in]{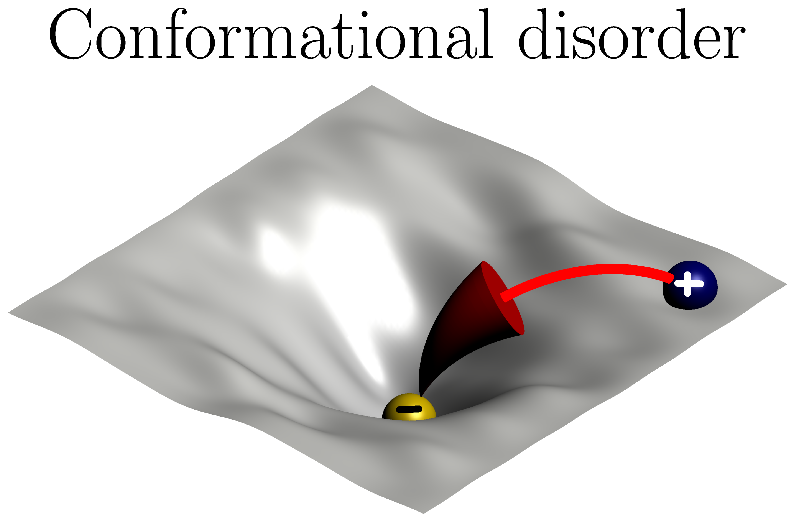}
\end{tocentry}

\begin{abstract}
We consider the bimolecular charge carrier recombination in amorphous organic semiconductors having a special kind of energetic disorder where energy levels for electrons and holes at a given transport site move in the same direction with the variation of some disorder governing parameter (the parallel disorder). This particular kind of  disorder could be found in materials where the dominant part of the energetic disorder is provided by the conformational disorder. Contrary to the recently studied case of electrostatic disorder, the conformational disorder, if spatially correlated,  leads to the increase of the recombination rate constant which becomes greater than the corresponding Langevin rate constant. Probably, organic semiconductors with the dominating conformational disorder represent the first class of amorphous organic semiconductors where the recombination rate constant could overcome the Langevin limit.
\end{abstract}


\newpage

\section{Introduction}

Charge carrier transport in amorphous organic semiconductors, apart from the exceptional case of pure monopolar transport, is inevitably accompanied by the carrier recombination. Recombination process could take different forms, here we consider the most universal case of bimolecular recombination which for the spatially homogeneous distribution of carriers is described by the second-order kinetics equation
\begin{equation}\label{kin}
    \frac{d(n,p)}{dt}=-\gamma np
\end{equation}
where $n(t)$, $p(t)$ are the concentration of electrons and holes, correspondingly, and $\gamma$ is the rate constant (we assume that the intrinsic concentration of carriers is negligible).

Recombination plays an ambiguous role in organic electronics. Firstly, it provides desired photons in organic light emitting diodes (OLEDs) but, secondly, it severely limits performance of organic photovoltaic (OPV) devices. For this reason the general trend in OPV is to develop mesoscopically inhomogeneous materials (e.g, as a fine mixture of electron and hole transporting materials) which provides separate pathways for the carriers with the opposite signs and severely diminishes recombination \cite{Lakhwani:557,Proctor:1941,Kniepert:1301401,Pivrikas:176806,Juska:2858,Deibel:163303,Deibel:075203}. Contrary to that, in OLEDs it is highly desirable to use materials and structures where the recombination is not strongly suppressed or, even better, is enhanced.

In amorphous organic semiconductors used in OLEDs the recombination rate constant is frequently assumed to obey the Langevin relation
\begin{equation}\label{Lan}
\gamma_{\rm L}=\frac{4\pi e}{\varepsilon}\left(\mu_{+}+\mu_{-}\right)
\end{equation}
where $\varepsilon$ is the dielectric constant of the medium and $\mu_{+}$, $\mu_{-}$ are mobilities of holes and electrons, correspondingly. \cite{Langevin:433} \Eq{Lan} was derived by Paul Langevin very long ago without any consideration of the disorder effects.

The very possibility of $\gamma$ to be approximately equal to $\gamma_{\rm L}$ in mesoscopically homogeneous amorphous semiconductors is still a matter of some controversy. Computer simulation\cite{Albrecht:455,Groves:155205,Holst:235202} as well as experimental data on OLEDs\cite{Blom:930,Blom:479,Dicker:45203,Pivrikas:125205,Kuik:4502,Kuik:093301,Wetzelaer:165204} suggest that the Langevin result still holds in homogeneous amorphous semiconductors.
At the same time, in our recent paper we suggested a simple model of the bimolecular recombination in amorphous organic semiconductors which takes into account strong spatial correlation of the random energy landscape inherent to such materials.\cite{Novikov:22856} Correlation naturally arises in amorphous semiconductors where the significant part of the total energetic disorder is provided by electrostatic contributions from randomly located and oriented permanent dipoles and quadrupoles. Our results clearly show that $\gamma$ in organic semiconductors having the correlated electrostatic disorder is smaller than the Langevin rate constant $\gamma_{\rm L}$, the so called reduction factor $\zeta =\gamma/\gamma_{\rm L}< 1$, even $\zeta \ll 1$ for strong disorder, and the earlier simulations gave an approximate equality $\gamma\approx\gamma_{\rm L}$ because the disorder per se does not break the Langevin relation. Uncorrelated disorder, considered in \onlinecites{Albrecht:455,Groves:155205,Holst:235202}, inevitably provides the Langevin rate constant irrespectively to the strength of the disorder. Even the simulation of the carrier recombination in phase-separated organic semiconductor blends, though not exactly adherent to our model of the homogeneous material, gives the rate constant close to the Langevin one if the domain size is small enough ($\le 5$ nm) and energetic disorder is spatially uncorrelated, again in full agreement with our conclusion \cite{Heiber:136602,Heiber:205204}.

Experimental data for $\zeta$ give, at the very best, the accuracy no better than one order of magnitude, so the conclusions in \onlinecites{Blom:930,Blom:479,Dicker:45203,Pivrikas:125205,Kuik:4502,Kuik:093301,Wetzelaer:165204} do not contradict our result. Besides, our results show that for low magnitude of the energetic disorder ($\sigma\simeq 0.05-0.07$ eV), typical for many devices, deviation of $\gamma$ from $\gamma_{\rm L}$ is not significant. Thus, assumption of $\gamma\approx \gamma_{\rm L}$ may give a reasonable description of some OLED parameters (see, for example, \onlinecites{Zhao:1181,Li:15154}). More detailed discussion is presented in \onlinecite{Novikov:22856}.

We carefully studied the recombination of charge carriers in polar amorphous organic materials where the dominant part of the energetic disorder is provided by permanent dipoles, though the most important conclusions are valid for other materials having electrostatic disorder, e.g. nonpolar organic materials with high concentration of permanent quadrupoles.

Hence, the major conclusion seems to be that the spatially correlated energetic disorder inevitably leads to the decrease of $\zeta$, and the correlation is the only necessary condition to observe such behavior. This conclusion is wrong. There is another necessary condition needed for the reduction of $\zeta$, namely, the specific construction of the electrostatic disorder. Essentially, this is the spatial disorder in the distribution of the electrostatic potential $\varphi(\vec{r})$, generated by randomly located and oriented dipoles and quadrupoles. Then $U(\vec{r})=e\varphi(\vec{r})$ becomes the random energy landscape for charge carriers where $e$ is the charge of the particular carrier, and charges of electrons and holes have the opposite signs. Hence, the inherent property of this kind of energetic disorder is the opposite landscapes for electrons and holes: a valley for electrons is a hill for holes. For this reason if, say, an electron is sitting at the bottom of a well and the hole is approaching, then it is attracted to the electron by the Coulomb force and at the same time is repulsed by the fluctuating potential which forms the well and is provided by dipoles and quadrupoles. Inevitably, this additional repulsion hinders the recombination and this is the reason for the decrease of $\gamma$ in comparison to $\gamma_{\rm L}$.

A natural question arises: is it possible to organize a correlated energetic disorder having the opposite property, i.e. the simbatic spatial variation of random energies of electrons and holes, where a valley for holes is a valley for electrons, and the same is true for the hills? Obviously, such disorder should lead to the additional attraction between electrons and holes and thus to the increase of the recombination rate constant in comparison to the Langevin value, i.e to $\zeta > 1$. We are going to demonstrate that this situation is indeed possible.

\section{Parallel disorder}

\begin{figure}[tbp]
\includegraphics[width=3.25in]{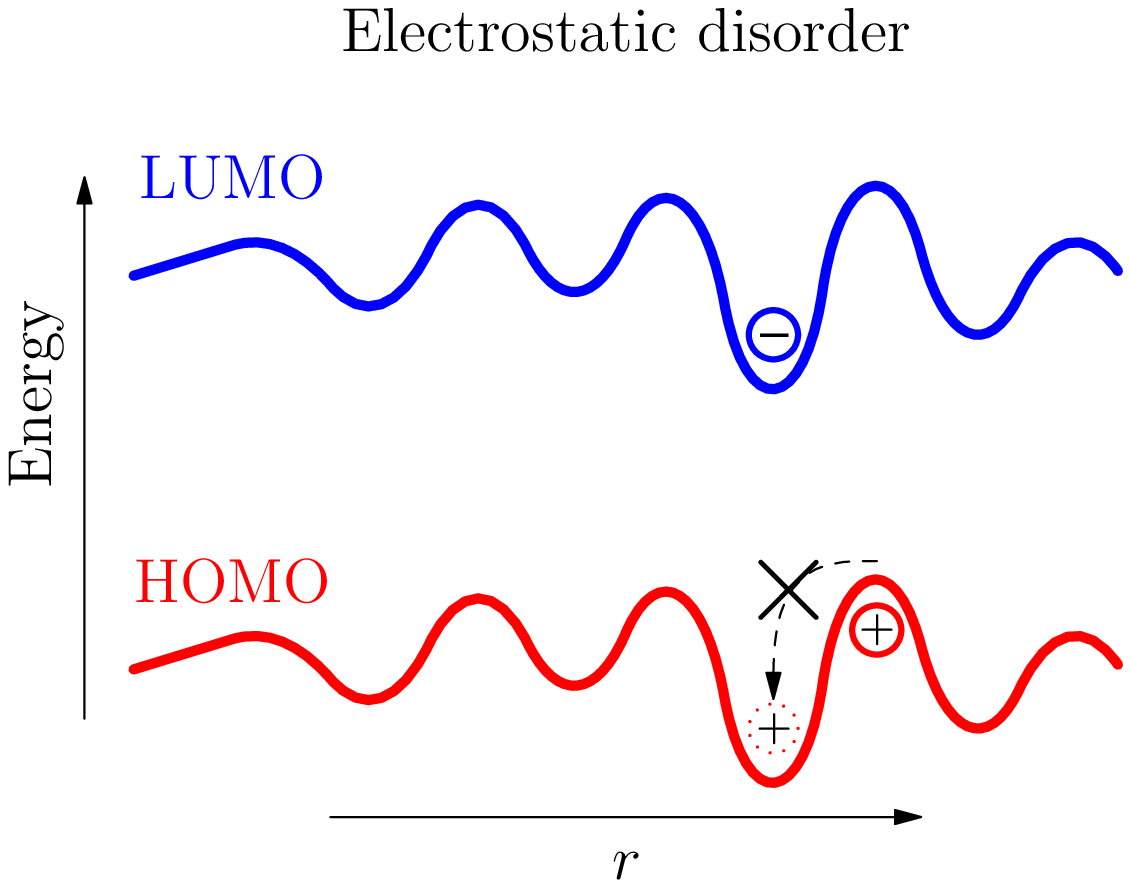}
\vskip10pt
a)

\medskip
\medskip
\includegraphics[width=3.25in]{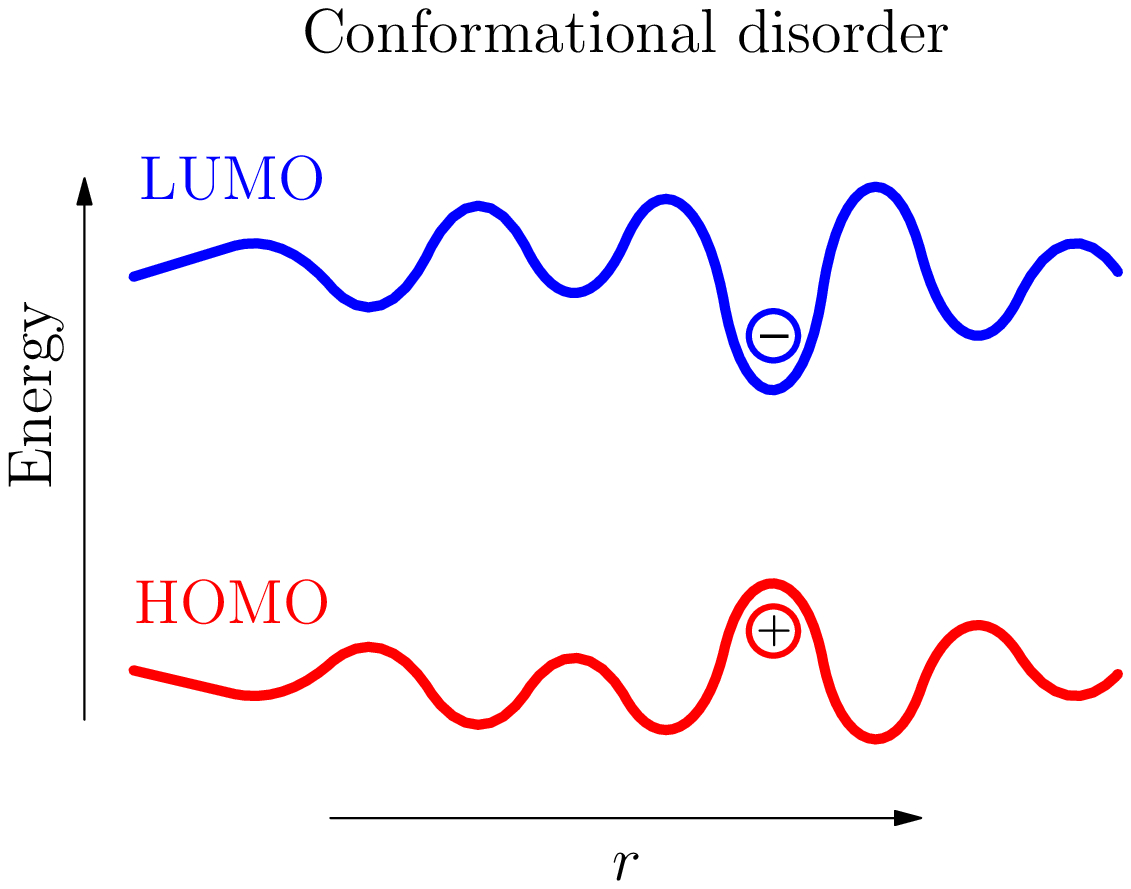}
\vskip10pt
b)
\caption{a) Spatial profiles of HOMO and LUMO energies of transport sites in amorphous organic semiconductor where the dominant part of the energetic disorder $U(\vec{r})$ is the electrostatic disorder. Crossed arrow indicates the obstacle for the spatial convergence of carriers due to the disorder. b) The same profiles for the conformational disorder (see explanation in the text).}  \label{gen-feature-el}
\end{figure}

The disorder we are seeking for should have the behavior shown in   \fig{gen-feature-el}b, the opposite one to the electrostatic disorder shown in  \fig{gen-feature-el}a. Mirrored spatial variation of the energies of HOMO and LUMO provides symbatic spatial variation of energies of electrons and holes. In fact, even electrostatic disorder could arrange the landscape having desired behavior, if the dominant part of disorder is provided by the polarizability mechanism \cite{Dunlap:80,May:136401}. In this case the random energy is proportional to the charge squared, so the energy profile is approximately identical for electrons and holes apart from the constant shift $\epsilon_{\rm LUMO}-\epsilon_{\rm HOMO}$. For this mechanism we should expect rather low rms disorder $\sigma\simeq 0.02 - 0.03$ eV \cite{Dunlap:80}, so the effect on recombination is not significant and, more important, this particular disorder could be easily swamped by other contributions not having the desired parallel property.

Recent paper \cite{Masse:115204} indicates that the desired energy landscape could be provided by the conformational disorder. The authors performed a careful \textit{ab initio} modelling of the structure of three amorphous organic semiconductors  $\alpha$-NPD, TCTA, and Spiro-DPVBi, popular for use as transport materials in organic electronic devices. They found that in $\alpha$-NPD and TCTA the total disorder is mostly of the electrostatic origin, the corresponding contribution of the electrostatic rms disorder $\sigma_{\rm Coul}$ to the total $\sigma$ is greater than the conformational contribution $\sigma_{\rm Conf}$. On the contrary, in Spiro-DPVBi the most important contribution to the fluctuations of energies $\epsilon_{\rm HOMO}$ and $\epsilon_{\rm LUMO}$ comes from the conformational contribution. These energies strongly depend on the dihedral angle $\alpha$ between fragments of Spiro-DPVBi molecule and variation of $\alpha$ typically moves $\epsilon_{\rm HOMO}(\alpha)$ and $\epsilon_{\rm LUMO}(\alpha)$ in the opposite directions. Scatter plots of $\epsilon_{\rm HOMO}$ vs $\epsilon_{\rm LUMO}$ for some particular sample of the amorphous Spiro-DPVBi demonstrate the Pearson product-moment correlation coefficient equals to $-0.45$ (i.e., it is negative, so the higher is HOMO the lower is LUMO), while for  $\alpha$-NPD and TCTA the corresponding coefficients are $0.4$ and $0.53$,  again demonstrating the dominance of the electrostatic disorder, where HOMO and LUMO levels move in the same direction. Hence, we should expect that the spatial variation of $\alpha$ for amorphous Spiro-DPVBi naturally provides the profile resembling presented in \fig{gen-feature-el}b, while for $\alpha$-NPD and TCTA the corresponding profiles qualitatively follow \fig{gen-feature-el}a. We are going to demonstrate that the behavior of the amorphous Spiro-DPVBi is not unique and should be expected in a wide class of materials with the dominance of the conformational disorder.

\begin{figure}[tbp]
\includegraphics[width=2.5in]{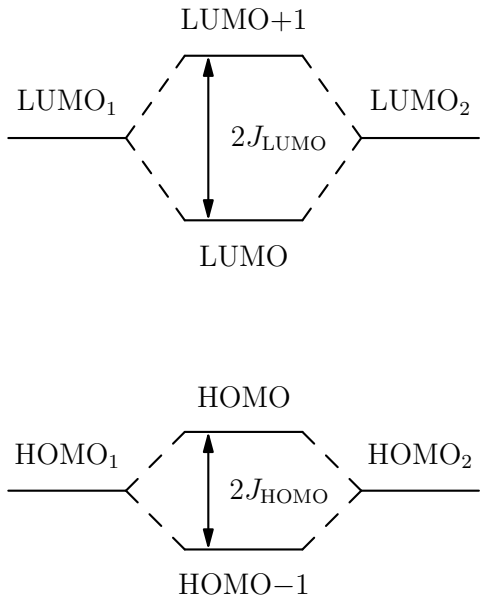}
\caption{Energy levels of the monomers (HOMO$_{1,2}$ and LUMO$_{1,2}$) and dimer molecule (in the middle).}
\label{levels}
\end{figure}

Let us consider a simple model of the formation of dimer molecule from identical aromatic entities. As the possible example we may use the difluorene molecule, the simplest member of the vast oligo- and polyfluorene family which provides useful charge transporting materials for organic electronics. We consider HOMO-LUMO pairs of both fragments and assume that inter-orbital interaction takes place exclusively in HOMO$_1$-HOMO$_2$ and LUMO$_1$-LUMO$_2$ pairs due to the large HOMO-LUMO gap in the monomer. Then the simplest ZINDO approach gives the approximate positions of the HOMO and LUMO in the dimer
\begin{eqnarray}\label{HOMO-LUMO}
&\epsilon_{\rm HOMO}=\epsilon_{\rm HOMO}^0+J_{\rm HOMO},\\
&\epsilon_{\rm LUMO}=\epsilon_{\rm LUMO}^0-J_{\rm LUMO},
\end{eqnarray}
where $\epsilon^0$ is the corresponding energy for monomer and $J_{\rm HOMO,LUMO} > 0$ are non-diagonal energy matrix elements, more detailed discussion could be found elsewhere\cite{Baumeier:11103}. Schematic picture of the energy levels arrangement is shown in \fig{levels}. Hence, if $J_{\rm HOMO}$ and $J_{\rm LUMO}$ varies in the same direction with the variation of some  disorder parameter in the material, we get precisely the proper mechanism for the realization of the parallel disorder. A natural mechanism for the variation of $J$ is the angle fluctuation in the dimer molecule, and the corresponding matrix elements could be approximated as $J(\alpha)\approx J_0 |\cos\alpha|$, where $\alpha$ is the dihedral angle between planes of almost flat fluorene units\cite{Troisi:4689}. We have the equilibrium angle $\alpha_{eq}\approx 45\degree$ (or $135\degree$) in amorphous oligofluorenes \cite{Chunwaschirasiri:107402}.

A notable effect for the recombination could be expected for $\sigma/kT \geq 1$. In Spiro-DPVBi for HOMO $\sigma_{\rm Conf}=0.103$ eV, $\sigma_{\rm tot}=0.122$ eV and for LUMO $\sigma_{\rm Conf}=0.146$ eV, $\sigma_{\rm tot}=0.156$ eV \cite{Masse:115204}. Assuming the difluorene-like structure and taking $J_0 \simeq 0.46$ eV, as suggested by Troisi and Shaw,\cite{Troisi:4689} we need fluctuations $\delta\alpha \simeq 8\degree-17\degree$ for $\sigma\simeq 0.05-0.1$ eV; such fluctuations look reasonable. We may assume also that for molecules having more complicated structure in comparison to difluorene even smaller angle fluctuations should be sufficient to provide a noticeable energetic disorder. Construction of the molecule from the non-identical entities does not provide any significant qualitative difference to our conclusions.

Summarizing, we see that the conformational disorder in the case of two connected aromatic fragments should provide quite naturally the landscape with properties required for the effective bimolecular recombination with enhancement (former reduction) factor $\zeta > 1$. The case of Spiro-DPVBi, though not exactly follows this model, could be considered as the first example of the diverse family of organic materials with the dominant conformational disorder.

\section{Analytical evaluation of the enhancement factor: effect of the spatial correlation}

Additional necessary condition for the enhanced recombination is the spatial correlation of the random energy landscape. In close analogy with the method developed in \onlinecite{Novikov:22856} and assuming the Gaussian density of states, typical for amorphous organic materials\cite{Bassler:15}, we may derive the general expression for the recombination rate constant $\gamma$.

We calculate the recombination rate constant for the pair of carriers with one of them sitting in the potential well with the depth $-U_0$ arranged by the random medium fluctuations and the carrier with the opposite charge approaching. The crucial approximation is the replacement of the true fluctuating potential energy $U(\vec{r})$ of the well by the conditionally averaged potential $-U_0 C(\vec{r})$ taking into account the condition $U(0)=-U_0$, here $C(\vec{r})=\left[\left<U(\vec{r})U(0)\right>-\left<U\right>^2\right]/\sigma^2$ is the random energy correlation function normalized in such a way that $C(0)=1$ and angular brackets mean the statistical average. Then we calculate the macroscopic rate constant $\gamma$ by averaging the pair rate constant over distribution of depths using the density of occupied states, see details of the calculation in \onlinecite{Novikov:22856}.

In fact, the only significant difference with the result of \onlinecite{Novikov:22856} is the inversion of the sign of the term $\propto U_0$ in $U_{\rm eff}(r)$ reflecting the transformation of the medium-induced additional repulsion to attraction
\begin{eqnarray}\label{gamma}
\gamma=\frac{4\pi D}{\left(2\pi\sigma^2\right)^{1/2}}
\int\limits_{-\infty}^\infty  \frac{dU_0}{\frac{4\pi D}{k_g}\exp\left[\beta  U_{\rm eff}(R)\right]+S(R)}
\exp\left[-\frac{(U_0-U_\sigma)^2}{2\sigma^2}\right] \\
S(r)=\int\limits_r^\infty \frac{dz}{z^2}\exp\left[\beta U_{\rm eff}(z)\right],\hskip10pt
U_{\rm eff}(r)=-\frac{e^2}{\varepsilon r}-U_0 C(r)
\end{eqnarray}
where $D=D_+ +D_-$ is the sum of the diffusivities of holes and electrons, $\beta=1/kT$, $U_\sigma=\beta\sigma^2$, $R$ is the radius of sphere where the recombination takes place, and $k_g$ is a rate constant of the quasi-geminate recombination occurring at short distances (in this paper we limit our consideration to the simplest case of the instant quasi-geminate recombination with $k_g\rightarrow\infty$). We assume also the spherical symmetry of the correlation function natural for the statistically isotropic and uniform random medium.

\begin{figure}[tbp]
\includegraphics[width=3in]{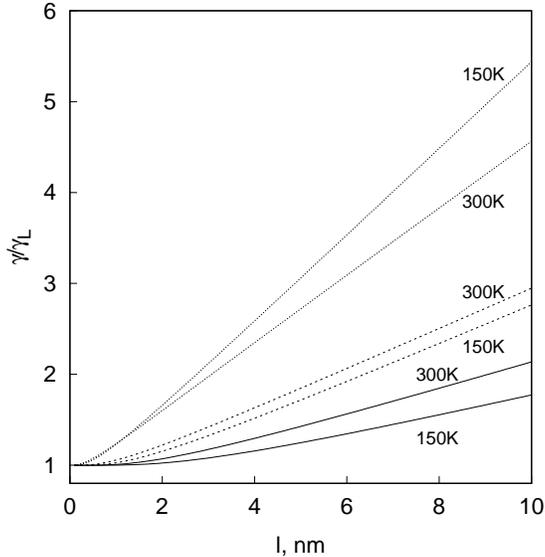}
\caption{Increase of the bimolecular recombination rate constant $\gamma$ in comparison to the Langevin rate constant $\gamma_{\rm L}$ according to \eq{gamma} for the case of instant  quasi-geminate recombination and stretched exponential correlation function \eq{C-exp} for two temperatures, 150K and 300K (indicated near the corresponding curves).  Solid, broken, and dotted lines show the behavior for $n$ equals to 1, 0.7, and 0.5, correspondingly. For other relevant parameters the typical values $R=1$ nm, $\varepsilon=3$, and $\sigma=0.1$ eV have been used. }  \label{exp-corr}
\end{figure}

In many realistic cases the correlation function of the random energy has the exponential form. Here we calculate $\gamma$ for a more general case of the stretched exponential correlation
\begin{equation}\label{C-exp}
C(r)=\exp\left[-\left(r/l\right)^n\right]
\end{equation}
where $l$ is the correlation length of the random energy landscape. General behavior of $\zeta$ for this correlation function is shown in \fig{exp-corr}.

We can provide the simple analytical estimation for $\zeta$ in the limit case $l\gg R$. For the typical situation we may assume that the lower limit in the integral for $S(R)$ is, in fact, equal to 0. Then the corresponding factor $S(R)$ does not depend on $R$ and could be written as the integral
\begin{eqnarray}\label{S-0}
S=\frac{1}{R_{\rm Ons}}\int\limits_0^\infty dp \exp\left[-G(p)\right]\\
G(p)= p+yy_\sigma\exp\left[-(p_0/p)^n\right]
\end{eqnarray}
where $y=U_0/\sigma$, $y_\sigma=\sigma/kT$, $p_0=R_{\rm Ons}/l$, and $R_{\rm Ons}=e^2/\varepsilon kT$ is the Onsager radius. Asymptotics of $G(p)$ are
\begin{equation}\label{G}
G(p)= \begin{cases} p, \hskip10pt p\ll p_0 \\
p+yy_\sigma, \hskip10pt p\gg p_0\end{cases}
\end{equation}
we assume here that $y> 0$ and $y\simeq y_\sigma$ (this means that we effectively carried out the integration over $U_0$). For the rough estimation of $S$ let us substitute the exact $G(p)$ with the asymptotic form \eq{G}, assuming a sharp transition between two asymptotics at $p=p_c$, and $p_c$ could be estimated from the relation
\begin{equation}\label{pc}
p_c\simeq y_\sigma^2\exp\left[-(p_0/p_c)^n\right]
\end{equation}
For $p_0/y_\sigma^2 \ll 1$ the leading asymptotics of the solution is
\begin{equation}\label{pc-sol}
p_c\simeq \frac{p_0}{\left[\ln\left(\frac{y_\sigma^2}{p_0}\right)\right]^{1/n}}
\end{equation}
and our approximation gives for $S$
\begin{equation}\label{G-est}
S\simeq \frac{1}{R_{\rm Ons}}\left[1-\exp(-p_c)+\exp(-y^2_\sigma-p_c)\right]
\end{equation}
Now if $p_c\ll 1$ but $p_c\gg \exp(-y_\sigma^2)$, then
\begin{equation}\label{gamma-est}
\gamma \simeq \frac{4\pi D R_{\rm Ons}}{p_c}
\end{equation}
and
\begin{equation}\label{zeta-est}
\zeta=\frac{\gamma}{\gamma_L}\simeq \frac{1}{p_c}\simeq\frac{l}{R_{\rm Ons}}\left\{\ln\left[\left(\frac{\sigma}{kT}\right)^2\frac{l}{R_{\rm Ons}}\right]\right\}^{1/n}
\end{equation}
(assuming the validity of the Einstein relation $\mu_\pm=eD_\pm/kT$). We see that $\zeta$ grows approximately linearly with $l$ for any $n$, apart from the insignificant logarithmic factor, in good agreement with \fig{exp-corr}, and the slope is in good agreement with the numerical calculation, too. Moreover, \eq{zeta-est} gives
\begin{equation}\label{d-zeta-est}
\frac{\partial \zeta}{\partial T}\simeq \frac{\zeta}{T}\left[1-\frac{1}{n\ln(T_0/T)}\right]
\end{equation}
where $T_0=l\varepsilon\sigma^2/ke^2$ (for $\sigma=0.1$ eV, $\varepsilon=3$, and $l=3$ nm $T_0=725$K), and we see that while $T < T_0$ there is a tendency of the derivative becomes negative with the decrease of $n$, again in the full agreement with \fig{exp-corr}.

\begin{figure}[tbp]
\includegraphics[width=3in]{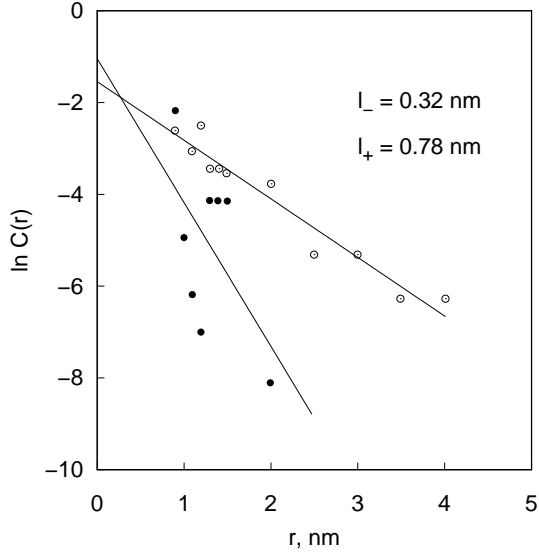}
\caption{Random energy correlation functions for electrons (solid dots) and holes (empty circles) for amorphous Spiro-DPVBi, data are borrowed from Figure 2 in \onlinecite{Masse:115204}. Straight lines show best fits for the equation $C(r)\propto\exp(-r/l)$; correlation lengths for electrons and holes are shown.}
\label{bobbert-eh}
\end{figure}

Simulation of the amorphous Spiro-DPVBi gives a spatially correlated energy landscape, having approximately exponential correlation functions for HOMO and LUMO (\fig{bobbert-eh}). Correlation lengths for electrons and holes are different. Probably, the significant part of the difference could be attributed to a very limited accuracy of the data, especially for the electrons. These lengths are comparable to the nearest neighbor distance in Spiro-DPVBi and, probably, do not lead to the noticeable increase of $\zeta$. Reliable experimental information on $\gamma$ in amorphous Spiro-DPVBi is still absent, device simulation is usually based on the assumption of the Langevin rate constant \cite{Ruhstaller:723}.

May we expect a larger correlation length in some amorphous organic semiconductors? If yes, what kind of organic semiconductors with dominated conformational disorder is favorable for the development of the correlated landscape?

\section{Which amorphous organic semiconductors are favorable for the development of the enhanced bimolecular recombination?}

Two necessary conditions for the realization of the enhancement mechanism in mesoscopically homogeneous amorphous semiconductors are: first, the proper type of the disorder, i.e. the parallel disorder, and, second, sufficiently strong spatial correlations.

As we see, in the particular case of the amorphous Spiro-DPVBi the proper type of the energetic landscape is provided by the conformational mechanism and our consideration indicates that the mechanism should be a common one. Nonetheless, by no means we may assume that the conformational disorder in all cases provides the proper parallel structure of the energy landscape. Yet we may  certainly suggest that the electrostatic contribution to the energetic disorder must be excluded as fully as possible. The simplest way to achieve such goal is to avoid atoms other than C and H in the molecule of the transport material. This is exactly the case for Spiro-DPVBi, its molecule contains neither N nor O atoms, and no halogen atoms as well. At the very least, the number of such atoms should be kept minimal. Spacious structure, again resembling the structure of Spiro-DPVBi, is favorable for the realisation of the high conformational $\sigma$.

The very possibility to achieve large correlation length is more problematic. As we already noted, in Spiro-DPVBi the correlation length, according to the simulation, is probably not large enough to provide a noticeable increase of $\zeta$.  It is natural to suggest liquid crystalline semiconductors \cite{Garnier:3334,McCulloch:328,Bushby:2012,Eichhorn:88,Woon:2311}  as promising materials for realization of high$-\zeta$ materials. Despite the very organized nature of liquid crystalline semiconductors it was found that the disorder effects are still very important.\cite{Duzhko:113312} Orientational correlation length in the glassy liquid crystalline materials (LCMs) below the glass temperature is rather large, we may expect $l \simeq 20-25$\AA \cite{Mansare:97} or even greater, up to few hundred angstroms \cite{prost1995physics}. As it was mentioned before, promising molecules for high$-\zeta$ semiconductors should contain the minimal amount of polar groups, absolutely none in the ideal case, and this is a very strict limitation for the typical LCMs. Nonetheless, fluorine containing LCMs could be suitable for the high-$\zeta$ semiconductors due to their low and highly isotropic polarity,\cite{Chen:2037,Dunmur:201,Dunmur:303,Achard:1387} and the true carbon-hydrogen LCMs do exist, though they are not very typical. \cite{Marrucci:10361} We may also mention the macrocyclic aromatic hydrocarbons, again containing only C and H atoms, demonstrating efficient bipolar charge transport, and perfectly capable to large scale ordering.\cite{Nakanishi:5435}

We briefly indicate here more exotic possibilities related to the formation of the amorphous organic semiconductors from the super-cooled liquids, where the existence of long-range correlations is commonly accepted, be it the so-called "Fischer clusters"\cite{Angell:3113,Angell:625} or correlation leading to the formation of the unusual amorphous "glacial phase"\cite{Alba-Simionesco:297,Hassaine:174508}; such correlations certainly should survive in the amorphous solid state. At the moment it is not clear how to manufacture organic semiconductor devices using  these materials and techniques.

\section{Conclusions}

We calculate the rate constant  of the bimolecular charge carrier recombination in amorphous organic semiconductors for the specific case of energetic disorder dominated by the conformational contribution. It was found previously that the conformational contribution is capable to organize the situation where HOMO and LUMO levels of transport molecules fluctuate in the opposite directions\cite{Masse:115204}, thus providing the disorder where an additional attraction between electrons and holes takes place, enhancing the recombination.

Consideration of the dimer model shows that the conformational parallel disorder should be quite common in amorphous semiconductors built of C and H atoms, and Spiro-DPVBi is, probably, just the first example of the large class of organic transport materials. We suggest that liquid crystalline organic semiconductors may be very promising materials for the formation of the correlated parallel disorder and emergence of the enhanced bimolecular recombination. Even modest increase in $\gamma$ with all other factors being the same could provide the comparable increase in the device efficiency for OLEDs. The importance of this possibility is difficult to overestimate.

\begin{acknowledgement}
Financial support from the Ministry of Science and Higher Education of the Russian Federation (A.N. Frumkin Institute) and Program of Basic Research of the National Research University Higher School of Economics is gratefully acknowledged.
\end{acknowledgement}

\bibstyle{achemso}
\bibliography{recombination-pd-R1}

\end{document}